\documentclass[twocolumn]{emulateapj}
\usepackage{natbib}
\usepackage{epsfig}
\usepackage{graphicx} 
\usepackage{subfigure}
\usepackage{float}
\usepackage{amsmath}
\usepackage{color}
\usepackage{amssymb}
\usepackage{amsfonts}
\usepackage{units} 
\usepackage{mathtools}
\usepackage{bm}
\usepackage[colorlinks,linkcolor=blue,anchorcolor=green,citecolor=blue]{hyperref}
\def\be{\begin{eqnarray}}
\def\ee{\end{eqnarray}}

\shorttitle{Relation between persistent emission luminosity and RM of FRBs}
\shortauthors{Yang, Li \& Zhang} 
\begin{document}

\title{Are persistent emission luminosity and rotation measure of fast radio bursts related?}

\author{Yuan-Pei Yang\altaffilmark{1}, Qiao-Chu Li\altaffilmark{2,3} and Bing Zhang\altaffilmark{4}} 

\affil{
$^1$ South-Western Institute for Astronomy Research, Yunnan University, Kunming, Yunnan, P.R.China; ypyang@ynu.edu.cn;\\
$^2$ School of Astronomy and Space Science, Nanjing University, Nanjing 210093, China;\\
$^3$ Key laboratory of Modern Astronomy and Astrophysics (Nanjing University), Ministry of Education, Nanjing 210093, China;\\
$^4$ Department of Physics and Astronomy, University of Nevada, Las Vegas, NV 89154, USA; zhang@physics.unlv.edu
}

\begin{abstract}

The physical origin of fast radio bursts (FRBs) is still unknown. Multiwavelength and polarization observations of an FRB source would be helpful to diagnose its progenitor and environment. So far only the first repeating source FRB 121102 appears to be spatially coincident with a persistent radio emission. Its bursts also have very large values of the Faraday rotation measure (RM) i.e., $|{\rm RM}|\sim10^5~{\rm rad~m^{-2}}$.  We show that theoretically there should be a simple relation between RM and the luminosity of the persistent source of an FRB source if the observed RM mostly arises from the persistent emission region. FRB 121102 follows this relation given that the magnetic field in the persistent emission region is highly ordered and that the number of relativistic electrons powering the persistent emission is comparable to that of non-relativistic electrons that contribute to RM. The non-detections of persistent emission sources from all other localized FRB sources are consistent with their relatively small RMs ($\left|{\rm RM}\right|\lesssim{\rm a~few}\times100~{\rm rad~m^{-2}}$) according to this relation. Based on this picture, the majority of FRBs without a large RM are not supposed to be associated with bright persistent sources.
\end{abstract} 

\keywords{radiation mechanisms: non-thermal --- radio continuum: general}

\section{Introduction}\label{intro}

Fast radio bursts (FRBs) are extragalactic radio transients with millisecond durations, large dispersion measures (DMs) and extremely high brightness temperatures \citep[e.g.,][]{lor07,tho13,cha17,chime19a,chime19b,ban19,rav19,pro19,mar20,fon20}.
The physical origin of FRBs is still unknown. 
Among all the published FRBs (http://frbcat.org, \citealt{pet16}), only the first repeating FRB source, FRB 121102, has a persistent radio counterpart and a large, evolving Faraday rotation measure (RM) \citep{cha17,mic18}.

FRB 121102 was first discovered with the Arecibo telescope \citep{spi14}. Its repeating behavior was further confirmed and studied with other radio telescopes all over the world, including Karl G. Jansky Very Large Array (VLA), Green Bank Telescope, the Five-hundred-meter Aperture Spherical Telescope (FAST), etc. \citep[e.g.][]{cha17,zyg18,di19}. 
Thanks to the precise localizations and multi-wavelength follow-up observations, the host galaxy of FRB 121102 was identified as a dwarf galaxy at redshift $z=0.193$ \citep{cha17,mar17,ten17}, and a persistent radio counterpart with luminosity of $\nu L_\nu\sim10^{39}~{\rm erg~s^{-1}}$ at a few GHz was discovered to be coincident with FRB 121102 spatially. 
The persistent emission has a non-thermal spectrum that deviates from a single power-law spectrum from $1~{\rm GHz}$ to $26~{\rm GHz}$ \citep{cha17}.
The RM of FRB 121102 has a very large absolute value, i.e., $|{\rm RM}|\sim 10^5~{\rm rad~m^{-2}}$, which decreased by ten percent during seven months \citep{mic18}. Such a large RM implies that the corresponding magnetic field is orders of magnitude stronger than that in the interstellar medium (ISM), and the variation RM might be related to the change of the magnetic field configuration \citep[e.g.,][]{zha18b} or strength  \citep[e.g.,][]{met19} along the line of sight. The host galaxy of another repeater, FRB 180916.J0158+65 (abbreviated as FRB 180916 as follows), was reported recently by \cite{mar20}, which is a
nearby massive spiral galaxy at $z=0.0337$. There is no coincident persistent emission above $3\sigma$ of $18~{\rm \mu Jy}$ at 1.6 GHz for this source, which places an upper limit on the persistent source luminosity $\nu L_\nu<7.6\times10^{35}~{\rm erg~s^{-1}}$. 
This upper limit is at least three orders of magnitude lower than that of FRB 121102, which gives one of the strongest constraints on the persistent emission luminosities of FRBs.

The persistent emission of FRB 121102 could be explained by the radiation from a nebula surrounding an FRB source, e.g. a supernova remnant (SNR) or a pulsar wind nebulae (PWN) \citep[e.g.,][]{yan16,mur16,met17,mar18}. Alternatively, it could be associated with a supermassive black hole \citep[e.g.][]{mic18,zha18b}. \citet{yan16} and \citet{li19} studied the synchrotron-heating process by an FRB source in a self-absorbed synchrotron nebula and found that the observed persistent emission associated with FRB 121102 could be generated via multi-injection of bursts. 
\citet{dai17} and \citet{yyh19} suggested that the persistent emission could be generated via an ultra-relativistic PWN sweeping up its ambient medium. \citet{wan19b} studied the multi-wavelength afterglow emission from the nebula powered by a repeating or non-repeating FRB central engine, and they found that a brighter nebula emission associated with a repeating source would have a larger RM.

In this work, we consider the possibility that the persistent emission and the RM of an FRB source originate from the same region. In this scenario, we derive a simple relation between the persistent emission luminosity and RM, and we find that FRB 121102 falls into the relation.
If this applies to all FRBs, our result implies that most FRBs, which have $\left|{\rm RM}\right|\lesssim{\rm a~few}\times100~{\rm rad~m^{-2}}$ \citep[e.g.,][]{pet17,bha18,cal18,osl19,ban19}, would not have detectable persistent emission with the current radio telescopes. This paper is organized as follows. In Section \ref{method}, we set up the theoretical framework for the relation between persistent emission and RM of an FRB source. We test the relation with some FRBs with the measurements of persistent emission and RM in Section \ref{data}. The results are summarized and discussed in Section \ref{summary}. 

\section{The $L_{\rm \nu,max}$-RM relation}\label{method}

Let us consider an FRB propagating in a plasma with number density of non-relativistic electrons $n_e$, magnetic field strength $B$, and scale length $R$. The RM in this region is given by
\be
\left|{\rm RM}\right|=\frac{e^3\xi_B B}{2\pi m_e^2c^4}n_e R,\label{RM}
\ee
where the parameter $\xi_B$ is defined as $\xi_B=\left<B_\parallel\right>/\left<B\right>$, $B_\parallel$ is the line-of-sight magnetic field, and the $\left<~\right>$ sign denotes the average value. For a random magnetic field, one has $\left<B_\parallel\right>=0$ and hence, $\xi_B=0$. Notice that in Eq.(\ref{RM}), we abbreviate $\left<B\right>$ to $B$.

On the other hand, the persistent emission with a continuous non-thermal spectrum is generally explained by synchrotron radiation from relativistic electrons. We assume that the number density of relativistic electrons in this region is $\zeta_en_e$, where $\zeta_e$ is the ratio between the relativistic and non-relativistic electron numbers\footnote{The RM contributed by relativistic electrons would be suppressed by a factor of $\gamma^2$, where $\gamma$ is the Lorentz factor of the electrons \citep{qua00}.}.
The radiation power and the characteristic synchrotron frequency from a randomly oriented electron with Lorentz factor $\gamma\gg1$ in a magnetic field $B$ are $P=(4/3)\sigma_{\rm T}c\gamma^2 B^2/8\pi$ and $\nu=\gamma^2eB/2\pi m_ec$, respectively.
Thus, the spectral radiation power is given by
$P_\nu\simeq P/\nu=m_ec^2\sigma_{\rm T}B/3e$, which is independent of $\gamma$.
Let the total number of relativistic electrons be $N_e=4\pi R^3\zeta_e n_e/3$. The maximum specific luminosity is
\be
L_{\nu,\max}&=&N_e P_\nu = \frac{64\pi^3}{27}m_ec^2\frac{\zeta_e}{\xi_B}R^2\left|{\rm RM}\right| \nonumber \\
& \simeq& (5.7\times10^{29}~{\rm erg~s^{-1}~Hz^{-1}}) 
\zeta_e\xi_B^{-1} \nonumber \\
&\times& \left(\frac{\left|{\rm RM}\right|}{10^5~{\rm rad~m^{-2}}}\right)\left(\frac{R}{0.01~{\rm pc}}\right)^2, \label{lum}
\ee
where  Eq.(\ref{RM}) has been used. One can see that there is a simple linear relation between $|{\rm RM}|$ and $L_{\rm \nu,max}$, with dependences on the size of the persistent emission regions and the parameters $\zeta_e$ and $\xi_B$.
The observed peak flux for an FRB source at the distance $D$ is
\be
F_{\nu,\max}&=&\frac{L_{\nu,\max}}{4\pi D^2}=\frac{16\pi^2}{27}m_ec^2\frac{\zeta_e}{\xi_B}\frac{R^2}{D^2}\left|{\rm RM}\right|
\nonumber \\
&\simeq & 480~{\rm \mu Jy}\zeta_e\xi_B^{-1} \left(\frac{\left|{\rm RM}\right|}{10^5~{\rm rad~m^{-2}}}\right) \nonumber \\
&\times&\left(\frac{R}{0.01~{\rm pc}}\right)^2\left(\frac{D}{1~{\rm Gpc}}\right)^{-2}.\label{fluxrm}
\ee 
For FRB 121102, the peak flux of the persistent emission is $\sim200~{\rm \mu Jy}$ \citep{cha17}, the RM is $\sim10^5~{\rm rad~m^{-2}}$ \citep{mic18}, and the persistent source has a projected size constrained to be $\lesssim0.7~{\rm pc}$ \citep{mar17}. The flux of the persistent emission source has a variation with a timescale of $\Delta t_{\rm per}\sim10~{\rm day}$ \citep[see Figure 2 of][]{cha17}, which further constrains the emission region to $R\sim c\Delta t_{\rm per}\simeq0.01~{\rm pc}$.
These numbers for FRB 121102 match Eq.(\ref{fluxrm}) very well 
given that $\xi_B\sim\zeta_e\sim1$ is satisfied.
This result might imply that the large RM of FRB 121102 is physically associated with its large persistent emission luminosity. 
The condition $\xi_B\sim1$ requires that the magnetic field is coherent in large scale, which is consistent with the large RM observation. The condition $\zeta_e\sim1$ requires that the number of relativistic and non-relativistic electrons are of the same order.
According to Eq.(\ref{fluxrm}), the variation of the persistent emission in the timescale of $\Delta t_{\rm per}\sim10~{\rm day}$ would result in an RM variation. \citet{mic18} reported that the RM of FRB 121102 decreases by $\sim10\%$ within seven months. Such a long-term evolution of the observed RM might be due to the change of the magnetic field configuration so that $\xi_B$ is a function of $t$ \citep[e.g.,][]{zha18b}. 

The above discussion assumes that both the magnetic field $B$ and the electron number density $n_e$ are uniform in a region with scale length $R$. If the magnetic field $B$ and electron number density $n_e$ satisfy a power-law distribution with radius from the source, the results should be of the same order of magnitude or somewhat lower compared with the uniform case presented in Eq.(\ref{fluxrm}).
A detailed discussion is presented in Appendix. 

For a source with the brightness temperature $T_{\rm B}$ and scale length $R$, the observed flux is
$F_\nu=(2kT_{\rm B}/\lambda^2) (R^2/D^2)$. According to Eq.(\ref{fluxrm}), one has
\be
\left|{\rm RM}\right|&=&\frac{27}{8\pi^2}\frac{\xi_B kT_{\rm B}}{\zeta_e\lambda^2m_ec^2} \simeq (0.6\times10^5~{\rm rad~m^{-2}}) \nonumber\\
&\times& \zeta_e^{-1}\xi_B\left(\frac{T_{\rm B}}{10^{12}~{\rm K}}\right)\left(\frac{\nu}{10~{\rm GHz}}\right)^2, \label{eq:Tb}
\ee
where we have normalized the frequency to 10 GHz, since the FRB 121102 persistent source has a broad spectrum extend beyond 10 GHz. In general, for an incoherent stationary source, the maximum brightness temperature is $T_{\rm B,max}\sim(10^{12}-10^{13})~{\rm K}$ due to the constraint of inverse Compton (IC) catastrophe \citep{kel69}. According to Eq.(\ref{eq:Tb}), the observed large RM from FRB 121102 demands a $T_b$ close to this limit. This provides direct evidence that the persistent emission associated with FRB 121102 originates from a strong compact radio source.

One should check the absorption effect in the emission region, including the Razin effect and free-free absorption. The synchrotron radiation of relativistic particles is subject to 
the plasma propagation effects. If the radiation frequency satisfies $\omega<\gamma\omega_p$, where $\gamma$ is the electron Lorentz factor, and $\omega_p=\sqrt{4\pi e^2 n_e/m_e}$ is the plasma frequency, the synchrotron spectrum would be cut off due to the suppression of beaming, which is called the Razin effect \citep[e.g.,][]{ryb79}. Using $\nu=\gamma^2 eB/2\pi m_ec$ and $\omega>\gamma\omega_p$, the condition for plasma transparency to the persistent emission is
\be
n_e<\frac{B\nu}{2ec}.
\label{razin}
\ee
According to Eq.(\ref{lum}) and Eq.(\ref{razin}), the transparency condition for the Razin effect can be written as
\be
n_e&<&3\times10^4~{\rm cm^{-3}}\zeta_e^{-1/2}\left(\frac{L_{\nu,\max}}{10^{29}~{\rm erg~s^{-1}Hz^{-1}}}\right)^{1/2}\nonumber\\
&\times&\left(\frac{\nu}{10~{\rm GHz}}\right)^{1/2}\left(\frac{R}{0.01~{\rm pc}}\right)^{-3/2}\\
&\simeq&6\times10^4~{\rm cm^{-3}}\xi_B^{-1/2}\left(\frac{|{\rm RM}|}{10^{5}~{\rm rad~m^{-2}}}\right)^{1/2}\nonumber\\
&\times&\left(\frac{\nu}{10~{\rm GHz}}\right)^{1/2}\left(\frac{R}{0.01~{\rm pc}}\right)^{-1/2}.
\ee
Thus, for FRB 121102, one needs $n<(10^4-10^5)~{\rm cm^{-3}}$.
On the other hand, we can also define the DM in the persistent emission source region as ${\rm DM_{src}}=n_e R$.
The above condition can be then also written as
\be
{\rm DM_{src}}
&\lesssim&655~{\rm pc~cm^{-3}}\xi_B^{-1/2}\nonumber\\
&\times&\left(\frac{\nu}{10~{\rm GHz}}\right)^{1/2}\left(\frac{|{\rm RM}|}{10^5~{\rm rad~m^{-2}}}\right)^{1/2}\left(\frac{R}{0.01~{\rm pc}}\right)^{1/2}.\nonumber\\
\label{eq:Razin}
\ee
Therefore, through measuring the rotation measure ${\rm RM}$, the persistent emission frequency $\nu$, and the variability timescale of the persistent emission $\Delta t_{\rm per}\sim R/c$, the upper limit DM of the emission region can be derived, which can be compared against that  the host galaxy DM constrained from the data.  We note that recent observations suggest that the excess DMs of several localized FRBs are generally consistent with being mostly due to the IGM contribution \citep{ban19,rav19,mar20}, suggesting that Eq.(\ref{eq:Razin}) is readily satisfied. Therefore, the Razin effect is not important unless the studied FRB has an abnormally large host DM.

Next, we consider the free-free absorption from the emission region. The free-free optical depth is given by
\be
\tau_{\rm ff}&=&\alpha_{\rm ff}R=0.018T^{-3/2}Z^2n_en_i\nu^{-2}\bar{g_{\rm ff}}R,
\ee
where $T$ is the thermal gas temperature, $n_e$ and $n_i$ are the number densities of electrons and ions, respectively, and $\bar{g_{\rm ff}}\sim1$ is Gaunt factor. Here $n_e=n_i$ and $Z=1$ are assumed for the emission region. The transparency condition requires $\tau_{\rm ff}<1$, which corresponds to 
\be
n_e<4\times10^5~{\rm cm^{-3}}\left(\frac{T}{10^4~{\rm K}}\right)^{3/4}\left(\frac{\nu}{10~{\rm GHz}}\right)\left(\frac{R}{0.01~{\rm pc}}\right)^{-1/2}.\nonumber\\
\label{eq:ff}
\ee
Combining the constraints from both Razin effect and free-free absorption, the transparency condition for the persistent emission of FRB 121102 requires $n_e\lesssim(10^4-10^5)~{\rm cm^{-3}}$. 
If the above conditions are not satisfied in some special environments (e.g. those FRBs with abnormally large source DM), the persistent emission luminosity would be much lower than that given by Eq.(\ref{lum}) due to the absorption effects.

\section{Persistent emission and rotation measure of fast radio bursts}\label{data}

\begin{table*}
\caption{ FRB sample with measurements (or upper limits) of RM and persistent emission flux \label{tab1}}
\begin{tabular}{ccccccccccc}
    \hline
    \hline
FRB Name  &  ${\rm DM_{obs}}^{\rm a}$   & $\rm{DM_{MW}}^{\rm b}$   & $z^{\rm c}$    &   $d_{\rm{L}}^{\rm d}$  & ${\rm RM}^{\rm e}$ & $F_{\nu,{\rm RM}}^{\rm f}$ & $F_{\nu}^{\rm g}$\ & $\nu^{\rm h}$ & $L_{\nu}^{\rm i}$  & References\\
  & ($\unit{pc\,cm^{-3}}$)   &  ($\unit{pc\,cm^{-3}}$)   &     &   ($\unit{Gpc}$)  &  ($\unit{rad\,m^{-2}}$)  & ($\mu\unit{Jy}$) & ($\mu\unit{Jy}$)  &  ($\unit{GHz}$) & ($10^{29}\unit{erg~s^{-1}Hz^{-1}}$)  &  \\
    \hline
FRB  121102 &  $557$       & $188 $ & $0.19273$    &   $0.98$  &  $1.4\times10^{5}$ & 698  &   180  & $1.7$ & $2.1$  & 1,2,3,4\\

FRB 180916  &  $348.76$ & $200$& $0.0337$&   $0.15$  & $-114.6$ & 24.4  & $<18$    & $1.6$  & $<0.0048$   & 5,6\\

FRB 180924  &  $361.42$ & $40.5$ & $0.3214$    &   $1.74$  & $14$ & 0.022         & $<20$     & $6.5$ & $<0.72$ & 7\\

FRB 181112	&  $589.27$ & $102$ & $0.47550$  & $2.76$ & $10.9$ & $0.0068$ & $<21$ & $6.5$ & $<1.91$ & 8\\

\hline

FRB 110523  &  $623.3$ & $43.52$ & $0.58^{+0.21}_{-0.21}$ & $3.5^{+1.6}_{-1.4}$ & $-186.1$ & $0.073$ & $<40$ & $0.8$ & $<5.8^{+6.6}_{-3.7}$ & 9\\

FRB 150215  &  $1105.6$  & $427.2$& $0.69^{+0.22}_{-0.22}$    &   $4.3^{+1.7}_{-1.6}$  & $1.5$ & $0.00039$    & $<6.48$   & $10.1$ & $<1.4^{+1.4}_{-0.9}$  & 10\\

FRB 150418	&	$776.2$ & $188.5$ & $0.59^{+0.21}_{-0.21}$ & $3.6^{+1.6}_{-1.4}$ & $36$ & $0.013$ & $<70$ & $1.4$ & $<11^{+12}_{-7}$ & 11\\

FRB 150807	&	$266.5$ & $36.9$ & $0.17^{+0.10}_{-0.11}$ & $0.85^{+0.57}_{-0.57}$ & $12$ & $0.08$ & $<240$ & $5.5$ & $<2.1^{+3.7}_{-1.8}$ & 12\\

FRB 160102  &  $2596.1$  & $13$   & $3.04^{+0.51}_{-0.48}$     &   $26.4^{+5.4}_{-4.9}$ & $-220.6$ & $0.0015$  & $<30$     & $5.9$ & $<249^{+112}_{-84}$   & 13,14\\

FRB 180309  &  $263.42$ & $44.69$& $0.16^{+0.10}_{-0.10}$    &   $0.79^{+0.57}_{-0.51}$  & $<150$   & $<1.2$       & $<105$     & $2.1$ & $<0.8^{+1.5}_{-0.7}$  & 15\\

FRB 191108	&  $588.1$ & $52$ & $0.53^{+0.20}_{-0.20}$ & $3.1^{+1.5}_{-1.3}$ & $474$ & $0.24$ & $<213$ & $1.4$ & $<24^{+29}_{-16}$ & 16\\
   \hline
   \hline	\\
   \end{tabular}
\footnotesize
$^{\mathrm{a}}$The observed DMs of FRBs.\\
$^{\mathrm{b}}$The DM contribution from Milky Way, which is from FRB catalog \citep{pet16}.\\
$^{\mathrm{c}}$The measured/inferred redshifts of FRBs. For FRB 121102, FRB 180916, FRB 180924 and FRB 181112, their redshifts are from the redshift measurements of their host galaxies \citep{cha17,ten17,ban19,mar20,pro19}. For other FRBs, their redshifts are inferred by the extragalactic DMs, i.e., Eq.(\ref{dme}). \\
$^{\mathrm{d}}$ The luminosity distance inferred by redshift.\\
$^{\mathrm{e}}$ The observed RMs of FRBs.\\
$^{\mathrm{f}}$ The predicted flux density given by RM, i.e., Eq.(\ref{fluxrm}).\\
$^{\mathrm{g}}$ The observed flux density of the persistent emission. For the FRBs without persistent emission detected, the upper limits correspond to the $3\sigma$ flux density limits.\\
$^{\mathrm{h}}$ The frequency at which the persistent emission is measured.\\
$^{\mathrm{i}}$ The persistent emission luminosity inferred by the observed flux density and luminosity distance.\\
References: (1) \cite{spi14}; (2) \cite{ten17}; (3) \cite{mar17}; (4) \cite{mic18}; (5) \cite{chi19}; (6) \cite{mar20}; (7) \cite{ban19}; (8) \cite{pro19}; (9) \cite{mas15};
(10) \cite{pet17}; (11) \cite{kea16}; (12) \cite{rav16}; (13) \cite{bha18}; (14) \cite{cal18}; (15) \cite{osl19}; (16) \cite{con20}.
\end{table*}

\begin{figure}[]
\centering
\includegraphics[angle=0,scale=0.42]{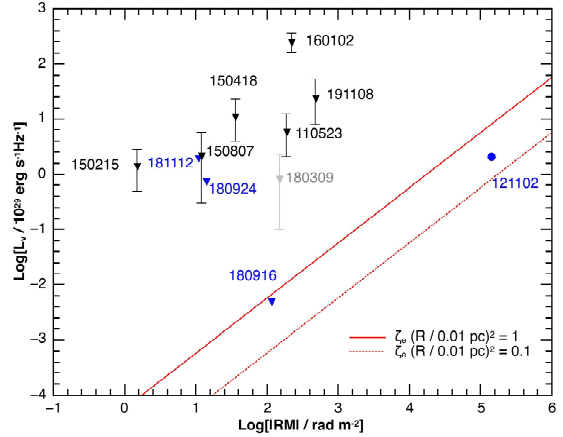}
\caption{The relation between specific luminosity of persistent emission and RM of FRBs. The red solid line and dotted line denote the predicted relation for $\zeta_e(R/0.01~{\rm pc})^2=1$ and $\zeta_e(R/0.01~{\rm pc})^2=0.1$, respectively. The black down-triangle points correspond to the FRBs with an upper limit of the persistent emission and a measured value of RM. The gray down-triangle point corresponds to the FRB with upper limits of RM and persistent emission. The circle point (FRB 121102) corresponds to the FRB with measured values of the persistent emission flux and RM. The blue points corresponds to the FRBs with precise localizations.}\label{fig1} 
\end{figure} 

In Table \ref{tab1}, we list ten FRBs from the FRB catalog of \citet{pet16}
with the measured RM and the measured values (or upper limits) of the persistent emission flux. Among them, only FRB 121102 has the measured values of both. 
For some FRBs, the persistent emission flux was constrained in a wide frequency range. We take the minimum value of the upper limits of the persistent emission flux density, because the predicted maximum flux density given by Eq.(\ref{fluxrm})  is independent of frequency. FRB 121102, FRB 180916, FRB 180924 and FRB 181112 have precise localizations \citep{cha17,ten17,ban19,mar20,pro19}, so the redshifts in Table \ref{tab1} are the directly measured values. For other FRBs, due to the lack of precise localization, we estimate their redshifts and luminosity distances via the extragalactic DM, i.e., ${\rm DM_E}={\rm DM_{obs}}-{\rm DM_{MW}}={\rm DM_{IGM}}+{\rm DM_{HG}}$, where ${\rm DM_{obs}}$ is the observed total DM, and ${\rm DM_{MW}}$, ${\rm DM_{IGM}}$, and ${\rm DM_{HG}}$ are the DMs contributed by Milky Way, intergalactic medium (IGM), and the FRB host galaxy, respectively. In this work, the DMs contributed by Milky Way are taken from\footnote{http://frbcat.org/} the FRB catalog \citep{pet16}, and we assume that the local DM contributed by FRB host galaxies is ${\rm DM_{HG,loc}}=100~{\rm pc~cm^{-3}}$ \citep[e.g.,][]{xu15,yan17b,luo18,li19b}. The extragalactic DM is given by \citep{den14} 
\begin{eqnarray}
\mathrm{DM}_{\rm E}(z)&=&\frac{3 c H_{0} \Omega_{b} f_{\mathrm{IGM}}}{8 \pi G m_{p}} \int_{0}^{z} \frac{\chi(z)(1+z) d z}{\left[\Omega_{m}(1+z)^{3}+\Omega_{\Lambda}\right]^{1 / 2}}\nonumber\\
&+&\frac{{\rm DM_{HG,loc}}}{1+z},\label{dme}
\end{eqnarray}
where the fraction of baryons in the IGM is $f_{\mathrm{IGM}} \sim 0.83$ \citep{fuk98,shu12}, and the free electron number per baryon in the universe is $\chi(z) \simeq 7 / 8$.
The $\Lambda$CDM cosmological parameters are taken as $\Omega_{\rm m}=0.315\,\pm\,0.007$, $\Omega_bh^2=0.02237\,\pm\,0.00015$, and $H_0=67.36\,\pm\,0.54\,\unit{km\,s^{-1}\,Mpc^{-1}}$ \citep{pla18}. 
For an FRB with redshift inferred from $\mathrm{DM}_{\rm E}$, the corresponding redshift error is given by $\sigma_{\rm IGM}$ due to the IGM density fluctuation \citep{mcq14}. For FRB 160102, since its redshift is out of the range given by \citet{mcq14}, we assume $\sigma_{\rm IGM}=350~{\rm pc~cm^{-3}}$.

In Figure \ref{fig1}, we plot the relation between the specific luminosity of the persistent emission and RM. The FRB data are taken from Table \ref{tab1}. 
According to Eq.(\ref{lum}), 
the red solid line corresponds to the predicted relation for $\zeta_e(R/0.01~{\rm pc})^2=1$, and the red dotted line corresponds to the predicted relation for $\zeta_e(R/0.01~{\rm pc})^2=0.1$.
For FRB 121102 with $|{\rm RM}|\sim10^5~{\rm rad~m^{-2}}$, the observed flux is closed to the predicted value for $\zeta_e(R/0.01~{\rm pc})^2\sim(0.1-1)$. 
For FRB 180916, the VLA observations shows that there is no coincident persistent emission above a $3\sigma$ rms noise level of $18~{\rm \mu Jy}$ per beam at 1.6 GHz \citep{mar20}. Such an upper limit is close to the predicted flux density given by Eq.(\ref{fluxrm}) for $\zeta_e(R/0.01~{\rm pc})^2\sim1$.
For other FRBs, the upper limits of the observed flux densities are significantly higher than the predicted persistent emission flux density.

\section{Discussion and Conclusion}\label{summary}

So far, both persistent radio emission and a large RM value are discovered only in FRB 121102. Although the persistent emission is found to be spatially coincident with the repeating bursts, it does not show a direct physical connection with the FRB 121102 bursting source \citep{cha17,mar17,ten17}.
In this work, we find  
a linear positive relation between the specific luminosity and RM (Eq.\ref{lum}). Such a relation is indeed satisfied for FRB 121102, given that the following conditions are satisfied:
\begin{itemize}
\item the persistent emission of FRB 121102 and its large RM originate from the same region; 
\item the magnetic field that contributes to RM and the persistent emission is coherent in large-scale (i.e. $\xi_B \sim 1$); 
\item the ratio between relativistic and non-relativistic electrons in the emission region, $\zeta_e$, is of the order of unity.
\item the Razin effect and free-free absorption are not significant, which requires $n_e\lesssim(10^4-10^5)~{\rm cm^{-3}}$ in the emission region for FRB 121102 and Eqs.(\ref{eq:Razin}) and (\ref{eq:ff}) in general.
\end{itemize}
If these conditions are satisfied for all other FRBs, we would expect that most FRBs with $\left|{\rm RM}\right|\lesssim{\rm a~few}\times100~{\rm rad~m^{-2}}$ \citep[e.g.,][]{pet17,bha18,cal18,osl19,ban19} would not have a detectable persistent emission counterpart. This seems to be consistent with the current observations. 
Considering that most FRBs have small RMs with $|{\rm RM}|\lesssim{\rm a~few}\times100~{\rm rad~m^{-2}}$, the expected luminosity of the persistent emission is $L_{\nu,{\rm RM}}\sim10^{27}~{\rm erg~s^{-1}Hz^{-1}}$ for $\zeta_e(R/0.01~{\rm pc})^2\sim1$. For a radio telescope with $3\sigma$ limiting fluxes of a few $\times10~{\rm \mu Jy}$, the observable distance for the persistent emission satisfies $d_{\rm L}\lesssim100~{\rm Mpc}$ or $z\lesssim0.03$. Thus, our model can be tested via the observations of nearby FRBs.
A deviation of the prediction Eq.(\ref{lum}) would suggest that at least one of the above conditions is not satisfied. For example, a bright persistent emission source with relatively small RM would suggest that the magnetic field in the persistent emission region is mostly random. At last, as shown in Figure \ref{fig1}, some FRBs (e.g., FRB 180916 and even FRB 110523) have an upper limit not too far away from the predicted zone (enclosed by the red lines). We suggest that observers may spend more observing times on target trying to make a positive detection of the persistent emission from these sources. A detection or a more stringent upper limit can help greatly to confirm or constrain the model proposed here.

The large-scale magnetic field requirement offers insight into the FRB mechanism. A large-scale magnetic field has been discovered near supermassive black holes or active galactic nuclei \citep{eat13,mic18}.  It was also hypothesized to exist in shocked nebula (e.g., SNR, PWN and etc.) surrounding a magnetized neutron star \citep[e.g.,][]{met19,mar18}. For the latter scenario, the synchrotron maser FRB mechanism requires that the magnetic fields lie in the plane of the shock. Such a field configuration needs to be destroyed to produce a radially ordered $B$ field in the region where RM is generated. 

As shown in Appendix, for more general setups, e.g.
$n_e\propto r^{-\alpha}$ and $B\propto r^{-\beta}$, for a given $\left|{\rm RM}\right|$, the predicted flux of persistent emission is of the same order or slightly lower than that given by Eq.(\ref{fluxrm}) for the uniform case. 
The observations of the persistent emission and RM of FRB 121102 imply that $\alpha+\beta<1$, which is close to the uniform distribution assumption. FRB 180916 \citep{mar20}, on the other hand, has an persistent emission flux upper limit very close to the predicted $L_{\rm \nu,max}-{\rm RM}$ relation, suggesting that either $\zeta_e<1$, or a smaller emission region ($R<0.01~{\rm pc}$), or a stratified medium with $\alpha+\beta>1$. 

Finally, different from DM measurements that require transients, RM measurements could be made for persistent sources as long as they are polarized. According to our analysis, the persistent emission region for FRB 121102 carries an ordered $B$ field, so that its emission should be linearly polarized. We suggest a direct measurement of RM of the persistent emission of FRB 121102 to test our prediction.

\acknowledgments 
We thank the anonymous referee for helpful comments and suggestions, and Qiancheng Liu and Xiaohui Sun for helpful discussions. This research has made use of the FRB catalog (http://frbcat.org, \citealt{pet16}).

\appendix

\section{Relation between persistent emission and rotation measure}
In this appendix, we perform a more general treatment on the relation between the persistent emission specific flux and the RM of FRBs.
We assume that at the radius $r_0<r<R$ from the FRB source, the electron number density follows $n_e=n_{e,0}(r/r_0)^{-\alpha}$ and the magnetic field follows $B=B_0(r/r_0)^{-\beta}$. Then the RM is given by
\be
\left|{\rm RM}\right|=\frac{e^3\xi_B}{2\pi m_e^2c^4}\int_{r_0}^R B(r) n_e(r) dr
=\left\{
\begin{aligned}
&\frac{e^3\xi_B}{2\pi m_e^2c^4}B_0n_{e,0}R\left(\frac{R}{r_0}\right)^{-(\alpha+\beta)},&~{\rm for}~\alpha+\beta<1,\\
&\frac{e^3\xi_B}{2\pi m_e^2c^4}B_0n_{e,0}r_0,&~{\rm for}~\alpha+\beta>1.
\end{aligned}
\right.\label{RM1}
\ee
In the radius range $r\sim r+dr$, the radiation power of a single electron is $P_\nu(r)=m_ec^2\sigma_{\rm T}B(r)/3e$, and the number of electrons is $4\pi r^2\zeta_en_e(r)dr$. The observed peak flux at the distance $D$ from the source is
\be
F_{\nu,\max}=\frac{1}{4\pi D^2}\int_{r_0}^RP_\nu(r) 4\pi r^2\zeta_en_e(r)dr=\frac{m_ec^2\sigma_{\rm T}\zeta_eB_0n_{e,0}}{3eD^2}\times\left\{
\begin{aligned}
&\frac{R^3}{3-(\alpha+\beta)}\left(\frac{R}{r_0}\right)^{-(\alpha+\beta)},&~{\rm for}~\alpha+\beta<3,\\
&\frac{r_0^3}{3-(\alpha+\beta)},&~{\rm for}~\alpha+\beta>3.
\end{aligned}
\right.\label{flux1}
\ee
According to Eq.(\ref{RM1}) and Eq.(\ref{flux1}), one finally has
\be
F_{\nu,\max}=\frac{16\pi^2}{9(3-\alpha-\beta)}m_ec^2\frac{\zeta_e}{\xi_B}\frac{R^2}{D^2}\left|{\rm RM}\right|\times\left\{
\begin{aligned}
&1,&{\rm for}~&\alpha+\beta<1,\\
&\left(\frac{R}{r_0}\right)^{1-(\alpha+\beta)},&{\rm for}~&1<\alpha+\beta<3,\\
&\left(\frac{R}{r_0}\right)^{-2},&{\rm for}~&\alpha+\beta>3.\\
\end{aligned}
\right.\label{flux1} 
\ee
This result is consistent with the uniform case with $\alpha=\beta=0$.
For any value of $\alpha+\beta$, one always has
\be
F_{\nu,\max}\leqslant\frac{16\pi^2}{9(3-\alpha-\beta)}m_ec^2\frac{\zeta_e}{\xi_B}\frac{R^2}{D^2}\left|{\rm RM}\right|. 
\ee
The equal sign corresponds to the case with $\alpha+\beta<1$.

\end{document}